\documentclass[12pt]{iopart}

\usepackage{graphicx}
\usepackage{dcolumn}
\usepackage{bm}

\begin{document}

\title{Neural networks using two-component Bose-Einstein condensates}

\author{Tim Byrnes$^{1}$, Shinsuke Koyama$^{2}$, Kai Yan$^1$, Yoshihisa Yamamoto$^{1,3}$}

\address{$^1$ National Institute of Informatics, 2-1-2
Hitotsubashi, Chiyoda-ku, Tokyo 101-8430, Japan}
\address{$^2$ The Institute of Statistical Mathematics, 10-3 Midori-cho, Tachikawa, Tokyo 190-8562, Japan } 
\address{$^3$ E. L. Ginzton Laboratory, Stanford University, Stanford, CA 94305}

\begin{abstract}
The authors previously considered a method solving optimization problems by using a system of 
interconnected network of two component Bose-Einstein condensates (Byrnes, Yan, Yamamoto New J. Phys. {\bf 13}, 113025 (2011)).  The use of bosonic particles was found to give a reduced time proportional to the number of bosons $ N $ for solving Ising model Hamiltonians by taking advantage of enhanced bosonic cooling rates.  In this paper we consider the same system in terms of neural networks.  We find that up to the accelerated cooling of the bosons the previously proposed system is equivalent to a stochastic continuous Hopfield network.  This makes it clear that the BEC network is a physical realization of a simulated annealing algorithm, with an additional speedup due to bosonic enhancement.  We discuss the BEC network in terms of typical neural network tasks such as learning and pattern recognition and find that the latter process may be accelerated by a factor of $ N $. 
\end{abstract}

\maketitle

\section{Introduction}

With the exception of a small class of problems that are
solvable analytically, most quantum many problems can only be examined using 
numerical means, for which exact simulations scale exponentially with the problem 
size.  Approximate methods such as quantum Monte Carlo and Density Matrix Renomalization
Group (DMRG) give accurate results for certain cases but no general algorithm exists that can be 
applied to an arbitrary system. In the field
of quantum information technology, quantum simulation has gathered a large amount of 
attention as an alternative means to study such 
quantum many body problems. A quantum simulator 
is a device where a quantum many body problem of interest
is artificially created in the lab, such that its properties can be controlled 
and measured directly \cite{buluta09}. Via directly using 
quantum mechanics in the simulation, there is no exponential overhead in keeping track of the 
number of states in the Hilbert space of the problem. This is in the spirit of 
Feynman's original motivations
for quantum computing \cite{feynman82}, where quantum mechanics, rather than classical mechanics, 
is used to simulate quantum many body problems. 

Given this general approach to quantum many problems, the question of whether a quantum 
simulation approach can be applied to a general Ising model becomes an important question.  
The Ising model problem consists of finding the lowest energy state of the Hamiltonian
\begin{equation}
\label{originalproblemham}
H_P = \sum_{i,j=1}^M J_{ij} \sigma_i \sigma_j + \sum_{i=1}^M \lambda_{i} \sigma_i,
\end{equation}
where $ J_{ij} $ is a real symmetric matrix that specifies the connections between the sites $i,j$, and $\sigma_i = \pm 1 $ is a spin variable. The task is then to find the minimal energy spin configuration $\{ \sigma_i \} $ of the Hamiltonian (\ref{originalproblemham}).    The problem of finding the solution of the Hamiltonian (\ref{originalproblemham}) is in fact known to be NP-complete, since it can be trivially be mapped onto the MAX-CUT problem \cite{mezard87}.  Furthermore, it can in principle encode an arbitrary NP-complete problem by a polynomial time mapping procedure \cite{ausiello99}, thus the potential application of a reliable method of solving (\ref{originalproblemham}) is extremely broad. Although (\ref{originalproblemham}) is itself a classical Hamiltonian since
it does not contain any non-commuting operators, as with quantum annealing where the Hamiltonian is supplemented with 
an additional transverse field, quantum ``tricks'' may be used to speed up the solution of the ground state beyond classical
methods. 

In a previous work we investigated a computational device which finds the solution of an Ising model
by a set of interconnected Bose-Einstein condensates \cite{byrnes11,yan11}.  In the approach of Ref. \cite{byrnes11}, each spin was replaced by a system of $ N $ bosons which can take one of two states. By implementing an analogous Hamiltonian to (\ref{originalproblemham})
and cooling the system down into into ground state, it was shown that the solution of the original Ising model problem could be found.  
There is a speedup compared to simply implementing (\ref{originalproblemham}) using single spins, because of the 
presence of bosonic final state stimulation within each bosonic spin. 
This resulted in finding the solution of (\ref{originalproblemham}) with a 
speedup of approximately $ N $. The attractive feature of the proposal in Ref. \cite{byrnes11} is that the computation 
is done simply by implementing a static Hamiltonian, so that no complicated gate sequence needs to be employed in order to 
find the ground state solution.  Effectively, the dissipative process itself performs the computation itself, and
therefore can also be considered to be a reservoir engineering approach to quantum computation. Related proposals were offered in Refs. \cite{utsunomiya11,takata12}, where instead of BECs, photons were used.

In this paper, we analyze the proposal in Ref. \cite{byrnes11} from the point of view of neural networks, specifically 
the stochastic continuous Hopfield model. Recasting the proposal in this form allows for a clearer analysis of the properties of the device, 
where standard results can be carried over to the BEC case.  It clarifies the origin of the $ \sim N $ speedup of the device, which was established via a numerical approach in Ref. \cite{byrnes11}. We find that the 
$ \sim N $ speedup originates from each element of the Hopfield network being accelerated due to bosonic stimulation, and thermal 
fluctuations provide the stochastic aspect to the Hopfield network. We then consider some 
simple applications of the BEC network for neural networking tasks.

\section{Bose-Einstein condensate networks}
\label{secdevice}

\begin{figure}
\scalebox{0.7}{\includegraphics{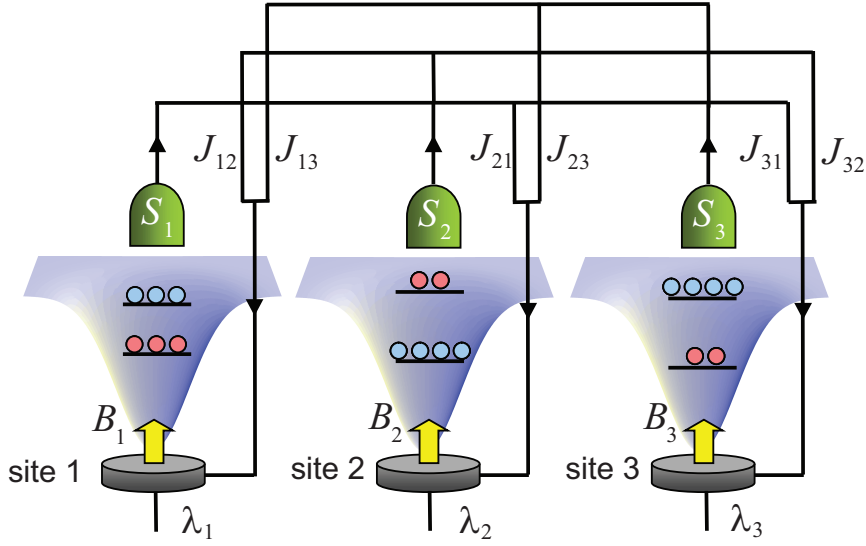}}
\caption{\label{fig1}
Each site of the Ising Hamiltonian is encoded as a trapping site, containing $N$  bosons. The bosons can occupy one of two states $\sigma=\pm1$, depicted as either red or blue. 
The interaction between the sites may be externally induced by measuring the total spin $ S_i $ on each site $i$ via the detectors.  A local field on each site equal to $B_i = \sum_j J_{ij} S_j + \lambda_i $ is applied via the feedback circuit. }
\end{figure}

In order to make this paper self-contained, we first give a brief description of the proposal of Ref. \cite{byrnes11}.  
In order to solve (\ref{originalproblemham}) we consider a system such as that shown in Figure \ref{fig1}. Each spin $\sigma_i $ in $ H_P $ is associated with a trapping site containing $N$  bosonic particles. The bosons can occupy one of two spin states, which we label by $\sigma=\pm1$. 
\begin{equation}
\label{kstates}
| k \rangle = \frac{1}{\sqrt{k(N-k)}} (a_{i+}^\dagger)^k (a_{i-}^\dagger)^{N-k} | 0 \rangle
\end{equation}
where $ a_{i\sigma} $ is the annihilation operator for spin $ \sigma $ on site $ i $.  On each site $ i $ we may define a total spin operator $ S_i = a_{i+}^\dagger a_{i+} - a_{i-}^\dagger a_{i-} $ taking eigenvalues $S_i | k \rangle = (-N+2k) | k \rangle $. 
The sites are controlled such that the system follows the bosonic Ising Hamiltonian
\begin{equation}
\label{eq2}
H = \sum_{i,j=1}^M J_{ij} S_i S_j + \sum_{i=1}^M \lambda_{i} S_i
\end{equation}
where  $J_{ij} $ is the same matrix as in $ H_P $ which specifies the computational problem. One such implementation of creating the 
interactions is via a measurement-feedback approach. In this approach, the total spin $ S_i $ on each site is continuously measured, and an appropriate control signal is continuously fed back into the system by applying a local dc field on another site. Given a measurement result of $ \{ S_j(t) \} $ across the spins, a local field 
\begin{equation}
\label{magneticfield}
B_i(t) = \sum_j J_{ij} S_j(t) + \lambda_{i}
\end{equation}
is applied on site $ i $.  Since the Zeeman energy due to this field is 
\begin{equation}
H = \sum_{i} B_i S_i ,
\end{equation}
a simple substitution 
yields (\ref{eq2}).

Starting from a random spin configuration $ \{ S_j(t) \} $, the system is cooled assuming that the ambient temperature $ T $ is fixed. 
The procedure is essentially identical to a simulated annealing procedure, the sole difference being the use of the bosonic Ising model. By varying the tempeature during the cooling process such that it is time dependent $ T(t) $, a strategy similar to thermal annealing may be performed, in order to escape being trapped in local minima.  In practice, instead of varying the temperature, 
varying the overall magnitude of (\ref{eq2}) by adjusting the strength of the magnetic field (\ref{magneticfield}) is equivalent. By 
taking advantange of the bosonic amplification of the cooling process, it was found in Ref. \cite{byrnes11} that an approximate factor
of $ N $ was found in the cooling process.

The time evolution is modeled by a extension of the method presented by Glauber (ref. 14) to bosons. Given the $ M  $ site Hamiltonian (\ref{eq2}), the states are labeled
\begin{equation}
| \bm{k} \rangle = \prod_{i=1}^M \frac{1}{\sqrt{k_i ! (N-k_i)! }} 
(a_{i+}^\dagger)^{k_i} (a_{i-}^\dagger)^{N-k_i} | 0 \rangle ,
\end{equation}
where the $ k_i $ range from 0 to $ N $, $ a_{i \sigma}^\dagger $ is the creation operator for a boson on site $ i $ in the state $ \sigma $, and we have defined the vector $ \bm{k} = (k_1, k_2, \dots , k_M) $. 
At $ t = 0 $, there is an equal probability $ p_{\bm{k}} $ of all states $ | \bm{k} \rangle $. The 
probability distribution then evolves according to 
\begin{equation}
\frac{d p_{\bm{k}}}{dt} = \sum_{i=1}^M 
- w( \bm{k}, \bm{\delta}_i) p_{\bm{k}} 
+ w( \bm{k}+\bm{\delta}_i, -\bm{\delta}_i) p_{\bm{k}+\bm{\delta}_i}
- w( \bm{k}, -\bm{\delta}_i) p_{\bm{k}} 
+ w( \bm{k}-\bm{\delta}_i, \bm{\delta}_i) p_{\bm{k}-\bm{\delta}_i} 
\label{twositerateequations}
\end{equation}
where $ \bm{\delta}_i $ is a vector of length $ M $ with its $i$th element equal to one and zero otherwise. The $ w( \bm{k}, \bm{\delta}_i) $ is a weight factor representing the transition $ | \bm{k} \rangle \rightarrow | \bm{k} + \bm{\delta}_i \rangle $,  containing both the bosonic final state stimulation factor and a
coefficient to ensure that the system evolves to the correct thermal equilibrium distribution.
We have restricted the transitions to first order transitions in (\ref{twositerateequations}) for simplicity. 
The final state stimulation factor
can be calculated by assuming a Hamiltonian $ H_t = \sum_{i \sigma} a_{i \sigma}^\dagger a_{i -\sigma} $ and 
calculating the transition rate according to Fermi's golden rule (up to a constant)
\begin{equation}
| \langle \bm{k}+\bm{\delta}_i | H_t | \bm{k} \rangle|^2 = (k_i+1)(N-k_i), 
\end{equation}
At thermal equilibrium, 
the transition rates are equal between the states $ | \bm{k} \rangle \leftrightarrow | \bm{k}+\bm{\delta}_i \rangle $, which ensures that $ \frac{d p_{\bm{k}}}{dt} = 0 $. The final state stimulation factors cancel for such pairs and we have
\begin{equation}
\frac{w (\bm{k},\bm{\delta}_i)}{w ( \bm{k}+\bm{\delta}_i, -\bm{\delta}_i)} = \frac{ p_{\bm{k}+\bm{\delta}_i} }{p_{\bm{k}}} ,
\end{equation}
and similarly for $ \bm{\delta}_i \rightarrow -\bm{\delta}_i$. From the probability distribution at thermal equilibrium we can calculate
\begin{equation}
\frac{ p_{\bm{k}+\bm{\delta}_i} }{p_{\bm{k}}} = \exp \left( 
-2 \beta \left[ \sum_{j\ne i}  J_{ij}( N  - 2 k_j) + \lambda_i \right]   \right) ,
\label{ratioexp}
\end{equation}
where $ \beta = \frac{1}{k_B T} $. This gives the coefficients as
\begin{eqnarray}
w(\bm{k}, \bm{\delta}_i) & = & \alpha (1 + \gamma_i) (k_i+1)(N-k_i)\nonumber \\
w(\bm{k}, -\bm{\delta}_i) & = & \alpha (1 - \gamma_i) k_i(N-k_i+1)
\end{eqnarray}
where
\begin{equation}
\gamma_i = \tanh \left(  -\beta \left[ \sum_{j\ne i} J_{ij}(  N  - 2 k_j) + \lambda_i \right] \right) ,
\end{equation}
$\alpha $ is a  constant determining the overall transition time scale. $ \alpha $ is set to 1 for all numerical calculations.

\section{Equivalence to the stochastic continuous Hopfield model}
\label{secequiv}
We now show that the scheme detailed in the previous section is formally equivalent to the stochastic continuous Hopfield model. After defining the Hopfield model we show the equivalences between the two systems, and derive the evolution equations for the BEC network in the context of the Hopfield model.

\subsection{Definition of the Hopfield model}

The Hopfield model is an asynchronous model of an artificial neural network \cite{rojas96,zurada92}, where the each unit of the network changes state at random times, such that no two units change simultaneously.  The Hopfield model is in the class of recurrent neural networks, where the output
of the network is fed back into itself, so that given a particular initial configuration, the system undergoes a sequence of changes in configuration of the units until steady state is achieved.  

In the standard (discrete, deterministic) Hopfield model, each unit takes the value $ \sigma_i = \pm 1 $.  The units are updated according to the rule
\begin{eqnarray}
\label{standardhopfield}
\sigma_i \rightarrow \sigma_i' = \mbox{sgn} \left[ \sum_j W_{ij} \sigma_j + b_i \right]
\end{eqnarray}
where $ \mbox{sgn} (x) = x/|x|  $ denotes the sign function, $ W_{ij} $ is a symmetric matrix with zero diagonal elements, $ b_i $ is the threshold of unit $ i $.  The units are updated to their new values $ \sigma_i' $ at random. Whether a given state is stable with respect to the updates can be determined by the Lyapunov (energy) function
\begin{equation}
E = \sum_{ij} W_{ij} \sigma_i \sigma_j + \sum_{i} b_i \sigma_i .
\end{equation}
The update sequence proceeds until a local minima with respect to the Lyanupov function is reached.  From a physics perspective, the transition rule (\ref{standardhopfield}) can be viewed as a cooling process at zero temperature, where given an initial high energy spin configuration, the spins cool one by one randomly into a low energy state. It is thus equivalent to the discrete Ising model problem (\ref{originalproblemham}) up to a contant energy factor originating from the diagonal elements of $ J_{ij} $, which play no part in the dynamics of the problem.  

The model can be extended to one where continuous variables $ x_i=[-1,1] $ are used in place of the discrete ones $ \sigma_i $. Such
continuous Hopfield networks have similar properties to the discrete version in terms of the configuration of stable
states \cite{hopfield84}. The way the model is usually defined is in terms of an electric circuit model (see for e.g. Ref. \cite{zurada92}).  
On a single unit $ i $, the time evolution of the circuit obeys
\begin{equation}
C_i \frac{d v_i}{dt} = -\frac{v_i}{R_i} + b_i + \sum_{j} W_{ij} x_i
\label{hopfielddyn}
\end{equation}
where $ C_i $ is the capacitance, $ R_i $ is the resistance, $ v_i $ is the voltage, and $ x_i $ is the output of the circuit after operation of the nonlinear amplifier (or activation function).  The activation function restricts the output to a limited region such that output is always in the range $ [-1,1] $. A typical choice is
\begin{equation}
x_i = \phi (v_i) = \tanh (v_i).
\end{equation}
The corresponding energy function for the dynamics is
\begin{equation}
E = \sum_{ij} W_{ij} x_i x_j + \sum_i b_i x_i + 
\sum_i \frac{1}{R_i} \int \phi^{-1} (x) dx  .
\end{equation}
From the energy function it may then be shown that the system is guaranteed to converge to a local minima of the 
energy.  Due to the energy structure of the continuous model being equivalent to the discrete model \cite{zurada92}, 
a solution to the continuous model then gives a one-to-one correspondence to the discrete model.

\subsection{Equations of motion on a single site}

In order to see the equivalence with the Hopfield model, let us first examine the dynamics on a single
site, and set $ M = 1$.  In this case the probability distribution of the states evolve as
\begin{eqnarray}
\frac{d p_{k}}{dt} = - w(k,1) p_{k} 
+ w(k+1,-1) p_{k+1} - w(k,-1) p_{k} 
+ w(k-1,1) p_{k-1}. \nonumber \\
\label{onerateequations}
\end{eqnarray}
where
\begin{eqnarray}
w(k,1) =  \alpha (1+ \gamma)(k+1)(N-k) \nonumber \\
w(k,-1)  =  \alpha (1- \gamma)k(N-k+1) \nonumber \\
\gamma = \tanh \left(  -\beta  \lambda \right) , \nonumber 
\end{eqnarray}
in this case since there is only one site. Multiplying the whole equation by $ k $ and summing over $ k $ given an equation for the mean value
\begin{equation}
\frac{d \langle k \rangle}{dt} = \alpha (1+ \gamma) \langle (k+1)(N-k) \rangle 
-\alpha (1- \gamma) \langle k(N-k+1) \rangle .
\end{equation}
Making the approximation that $ \langle k^2 \rangle \approx \langle k \rangle^2 $, and changing variables to $ S = -N + 2k $ the equation can be recast into the form
\begin{equation}
\frac{1}{\alpha} \frac{d s}{dt} = - N \gamma s^2 - 2 s
+  \gamma (2+ N)
\label{singlesiteboson}
\end{equation}
where we have used the normalized variable $ s = \langle S \rangle /N $.  An explicit solution for this may be found
to be
\begin{equation}
\label{onesiteevol}
s(t) = A \tanh \left(  \alpha \gamma A N t + K_0 \right) - \frac{1}{N \gamma}
\end{equation}
where $ A=\sqrt{1+\frac{2}{N} + \frac{1}{N^2 \gamma^2}} $ is a constant that is order unity at low temperatures such that $ N\gamma \gg 1 $.  $ K_0 $ is fixed by the initial conditions, for which we typically assume $ s(t=0) = 0 $.  
We see that the spin approaches its steady state value with a timescale $ \sim 1/\alpha |\gamma| N $.  At zero temperature where $ \gamma = \pm 1 $ (depending upon the direction of the applied external field), the time scale is enhanced by a factor of $ N $, which is due to bosonic final state stimulation, as found in Ref. \cite{byrnes11}.  

Let us compare this to the Hopfield network for a single site.  In this case the equation of motion reads
\begin{equation}
C \frac{d v}{dt} = -\frac{v}{R} + b.
\end{equation}
Changing variables to the output of the nonlinear amplifier, we obtain
\begin{equation}
C \frac{dx}{dt} = \left( \frac{d \phi^{-1}}{dx} \right)^{-1} \left[ - \frac{\phi^{-1} (x)}{R} +b \right].
\label{hopfieldonesite}
\end{equation}
This is the counterpart of the equation (\ref{singlesiteboson}) for the Hopfield case.  The solution of this
\begin{equation}
x(t) = \phi(bR + \exp[-t/CR+ K_0'] ).
\end{equation}
where $K_0' $ is set by the initial conditions.   

Writing the equations in this form makes the correspondence between the BEC system and the Hopfield network clear, which
we summarize in Table \ref{table1}. The output $ x $ of the Hopfield network corresponds to the normalized spin variable $ s $.   The magnetic field $ B $ applied on each site for the BEC system then corresponds to the voltage on each Hopfield unit before the non-linear amplifier.  The steady state values are determined by setting the right hand side of (\ref{singlesiteboson}) and (\ref{hopfieldonesite}) to zero.  The value of the steady state depends upon the ratio of the field ($ \lambda $ or $ b $) applied and the temperature $ k_B T = 1/\beta $ or the conductance $ 1/R $ respectively. 
The overall time constant is controlled by $ 1/\alpha $ or $ C $ in each case.  In the Hopfield network, there is obviously
no bosonic final state stimulation, so the speedup proportional to $ N $ is absent.  While the exact equation of motion for the two cases (\ref{singlesiteboson}) and (\ref{hopfieldonesite}) differ in their details, it is clear that the qualitatively there is a similar structure and behavior to the dynamics.  While in the Hopfield network, the overall time constant is determined by the capacitance of the circuit, the fundamental timescale of the BEC network is determined by the cooling rate of the bosons on each site.  

The analogue of the activation function $ \phi(v) $ may be derived as follows.  Considering the activation function 
as a rule that converts the internal magnetic field $ B $ to the spin variable $ s $, we may derive the average spin
at thermal equilibrium using the Boltzmann distribution taking into account of bosonic statistics.  From the partition function 
\begin{equation}
p_k^{\mbox{\tiny eq}} = (1-\exp(-2\beta B) ) \exp( - 2 \beta B k ),
\end{equation}
and therefore
\begin{eqnarray}
s^{\mbox{\tiny eq}}&=\sum_k p_k^{\mbox{\tiny eq}} (-1+2k/N)  \nonumber \\
&= \Phi(z) \equiv \frac{ (1- e^{2z(2+N)} ) + (2+N)(e^{2z(N+1)} - e^{2z} )}{(1-e^{2z})(1-e^{2z(1+N)})}
\label{equils}
\end{eqnarray}
where $ z = B \beta $ here.  The function above has a dependence that has similar behavior to 
\begin{equation}
\Phi(z) \approx \tanh \left( -\frac{z(N+2)}{3} \right) 
\end{equation}
which makes it clear that it plays a similar role to that of the activation function $ \phi $ in the Hopfield 
network.

\begin{table}
\begin{center}
\begin{tabular}{ccc}
\hline
Quantity & BEC network  & Hopfield model \\
\hline
Site variable & $s_i = \langle S_i \rangle/N $ & $ x_i $ \\
Site field variable & $ B_i $ &  $ v_i $ \\
Ising matrix & $ J_{ij} $ & $ W_{ij} $ \\
Local field & $ \lambda_i $ & $ b_i $ \\
Time constant & $ 1/\alpha $  & $ C $ \\
Steady-state control parameter & $ 1/k_B T $  & $ R_i $ \\
Activation function & $ \Phi $ & $ \phi $ \\ 
\hline
\end{tabular}
\caption{Equivalences between the network of BECs proposed in Ref. \cite{byrnes11} and the Hopfield model.}
\label{table1}
\end{center}
\end{table}

\subsection{Equations of motion for interconnected BECs}

We may now generalize to the multi-site case.  Multiplying (\ref{twositerateequations}) by $ k_i $ and summing over all $ \bm{k} $ gives the equations
\begin{equation}
\frac{d \langle k_i \rangle}{dt} = \alpha \langle  (1+ \gamma_i (\bm{k}) ) (k_i+1)(N-k_i) \rangle 
-\alpha \langle (1- \gamma_i(\bm{k}))  k_i(N-k_i+1) \rangle .
\end{equation}
where we have written $\gamma_i \rightarrow \gamma_i(\bm{k}) $ to remind ourselves that this is not a constant 
in this case. Making the approximation that $ \langle k^m \rangle \approx \langle k \rangle^m $, and making the change of 
variables to $ s_i = \langle S_i \rangle/N $, we obtain
\begin{equation}
\frac{1}{\alpha} \frac{d s_i}{dt} = - N \gamma_i (\bm{s}) s_i^2 - 2 s_i
+  \gamma_i (\bm{s}) (2+ N) .
\label{alleqns}
\end{equation}
The sole difference to the single site case here is that the equilibrium values 
of $ s_i $ are now dependent on the spins of all the other sites $ \bm{s} $. The dynamics
on each site is the same as the single site case, and thus evolves in time as (\ref{onesiteevol}),
considering the other spins to be approximately fixed. This basic structure is precisely the same
dynamics that determine the equation of motion of the Hopfield network (\ref{hopfielddyn}). 

Although it is not possible to solve
the set of equations (\ref{alleqns}) analytically, we may see in this formulation 
why the whole system should have a speedup of $ \sim N $, as found in Ref. \cite{byrnes11}. 
Considering a asynchronous update procedure (in fact this is exactly what was performed
to simulate the dynamics of the system in the Monte Carlo procedure \cite{yan11}), then all
but one ``active'' site is fixed in spins.  The active site then evolves in time according 
to the evolution of (\ref{onesiteevol}).  This has a speedup of $ \sim N | \gamma |$ in the 
evolution of the spin, thus to make an incremental change $ \delta s $ in the active spin
takes a time reduced by $N | \gamma | $ compared to the $ N = 1 $ case.  The spin is then fixed,
and then another site is chosen at random and this is updated.  Since each step takes
a reduced time of $ N | \gamma | $, the whole evolution proceeds at an aceelerated rate.  For 
sufficiently low temperatures, $ \gamma \approx \pm 1 $, and therefore the speedup 
is approximately $ \sim N $.

\subsection{Stochastic Hopfield network}

Up to this point, the equations of motion (\ref{alleqns}) have been entirely deterministic.  The role of the temperature was to merely shift the equilibrium values of the spins, as determined by (\ref{equils}), and did not contribute to any stochastic evolution of the system.  In the BEC network there are thermal fluctuations which cause the system to behave in a random manner.  Therefore in order to fully capture the dynamics of the BEC network we must include the contribution of the random thermal process.  Such stochastic versions of Hopfield networks, and their generalization to the Boltzmann machine (by having additional hidden units) are defined by modifying the update rule (\ref{standardhopfield}) to include probabilistic effects. Considering the discrete Hopfield model first, the algorithm consists of selecting a particular spin $ \sigma_i $, and making an update according to 
\begin{eqnarray}
\sigma_i & \rightarrow -\sigma_i  \hspace{2cm} & \mbox{if $\Delta E <0$} \nonumber \\
\sigma_i & \rightarrow -\sigma_i \hspace{2cm}  & \mbox{if $\Delta E >0$ with probability $ e^{-\Delta E/ k_B T} $} \nonumber \\
\sigma_i & \rightarrow \sigma_i \hspace{2cm}  & \mbox{if $\Delta E >0$ and otherwise} .
\end{eqnarray}
$ \Delta E $ is the energy difference between 
the state with the flipped spin and no flipped spin \cite{dudaxx}.  

The evolution equations (\ref{twositerateequations}) for the BEC network can be converted into a set of stochastic update rules which give 
the same time depedence when an ensemble average is taken.  The stochastic formulation also
allows for a convenient method of numerically simulating the system, which was discussed in detail in Ref. \cite{yan11}. 
We briefly describe the procedure as applied to the current formulation of the BEC network. 
The simulation is started from a random initial value of ${\bm k}=(k_1,k_2,...,k_M)$ in (\ref{twositerateequations}), and we update the system by repeating the stochastic transition process following the kinetic Monte Carlo method \cite{voter}. In each update we calculate the transition weight $w({\bm k},{\bm \delta}_i)$ in (\ref{twositerateequations}) for all the possible transitions. The transition is then made with a probability in proportion to the transition weight $w({\bm k}, {\bm \delta}_i)$. The time increment 
is then calculated according to 
\begin{eqnarray}
\Delta t = - \ln (r) /W_{\mbox{\tiny tot}}
\label{eq:dTincrement}
\end{eqnarray}
where $r \in (0,1]$ is a randomly generated number and 
\begin{equation}
W_{\mbox{\tiny tot}} = \sum_i ( w({\bm k},{\bm \delta}_i) + w({\bm k},-{\bm \delta}_i) )
\end{equation}
This procedure is repeated for many trajectories so that ensemble averages of quantities such as the average spin 
can be taken.  

This procedure produces exactly the same time dynamics as (\ref{twositerateequations}), hence for quantities such as the 
equilibration time this procedure must be followed.  However, if only the behavior at equilibrium is required, the update procedure 
can be replaced by the Metropolis algorithm.  The update procedure is then as follows. Start from a random initial value of ${\bm k}=(k_1,k_2,...,k_M)$.  Then make an update according to
\begin{eqnarray}
k_i & \rightarrow k_i \pm 1  \hspace{2cm} & \mbox{if $\Delta E <0$} \nonumber \\
k_i & \rightarrow k_i \pm 1 \hspace{2cm}  & \mbox{if $\Delta E >0$ with probability $ e^{-\Delta E/ k_B T} $} \nonumber \\
k_i & \rightarrow k_i  \hspace{2cm}  & \mbox{if $\Delta E >0$ and otherwise} .
\label{metropolisbec}
\end{eqnarray}
where the energy difference in this case is
\begin{equation}
\Delta E = \pm 2 \left[ \sum_{j\ne i}  J_{ij}( N  - 2 k_j) + \lambda_i \right]  .
\end{equation}
The exponential factor is precisely the same as that determining the weight factors in (\ref{ratioexp}).
The only difference in this case is that the bosonic stimulation factors are not present in this case, 
which are important only for determining the transition rates, and not the equilibrium values. The 
above Metropolis transition rule (\ref{ratioexp}) is identical to the stochastic Hopfield network, up to 
the difference that each site contains energy levels between $ k_i =0,\dots,N $.  It is therefore evidents 
in this context that the two systems are equivalent in their dynamics.   

We may also derive the equivalent stochastic differential equations for the average spin by adding 
noise terms to (\ref{alleqns}). First consider a single site, and start with the probability distribution 
(\ref{onerateequations}).   Assuming that $ N \gg 1 $, introduce the variables $ z = k/N $, $ \epsilon = 1/N $,
and the density $ q(z,t) = p_k / \epsilon $ of $ z $ at time $ t $, the master equation is rewritten
\begin{equation}
\frac{\partial q (z,t)}{\partial t} = - w(z,\epsilon) q(z,t) + w(z+\epsilon,-\epsilon) q(z+\epsilon,t)
- w(z,-\epsilon) q(z,t) + w(z-\epsilon),t)
\end{equation}
Expanding $ w(z \pm \epsilon,\mp \epsilon)$ and $ q(z \pm \epsilon, t ) $ up to second order in $ \epsilon $,
we obtain
\begin{eqnarray}
\frac{\partial q(z,t)}{\partial t} & = &
-\frac{\partial}{\partial z} \left[ (w(z,\epsilon)-w(z,\epsilon) ) \epsilon q(z,t) \right] \nonumber \\
& & + \frac{1}{2} \frac{\partial^2}{\partial z^2} \left[ ( w(z,\epsilon) + w(z,-\epsilon) ) \epsilon^2 q(z,t) \right] + O(\epsilon^3) \nonumber \\
& = & - \frac{\partial}{\partial z} \left[ A_{\epsilon} (z) q(z,t) \right] 
+ \frac{1}{2} \frac{\partial^2}{\partial z^2} \left[ B_{\epsilon} (z) q(z,t) \right] + O(\epsilon^3)
\end{eqnarray}
where
\begin{eqnarray}
A_{\epsilon} (z) & = &( w(z,\epsilon) -  w(z,-\epsilon))\epsilon  \nonumber \\
B_{\epsilon} (z) & = &( w(z,\epsilon) +  w(z,-\epsilon))\epsilon^2  .
\end{eqnarray}
Using the diffusion approximation such that the transition rates are on the order of $ w(z,\pm \epsilon) \sim 1/dt $, $ (w(z,\epsilon)-w(z,\epsilon)) \sim \epsilon/dt $, and $ \epsilon^2/dt \sim O(1) $, and taking the limits of $ \epsilon \rightarrow 0 $,
we obtain the Fokker-Planck equation
\begin{equation}
\frac{\partial q (z,t)}{\partial t} = - \frac{\partial}{\partial z} \left[ A(z) q(z,t) \right] 
+ \frac{1}{2} \frac{\partial^2}{\partial z^2} \left[ B(z) q(z,t) \right] 
\end{equation}
where $ A(z) = \lim_{\epsilon \rightarrow 0 } A_{\epsilon} (z) $ and $ B(z) = \lim_{\epsilon \rightarrow 0 } B_{\epsilon} (z) $.  The 
corresponding stochastic differential equation is given by
\begin{equation}
\frac{dz}{dt} = A(z) + \sqrt{B(z)} \xi(t)
\end{equation}
where $ \xi(t) $ is Gaussian white noise with $ \langle \xi(t) \rangle = 0 $ and $ \langle \xi(t) \xi(t') \rangle = \delta (t-t') $.  
Changing variables to $ s = -1+ 2z $, the coefficients for our case are
\begin{eqnarray}
A(z) &= & \frac{\alpha}{2} \left[ -N \gamma s^2 -2s + \gamma(2+N) \right]  \nonumber \\
B(z) & = & \frac{\alpha}{2} \left[ (1+s)(1-s) + \frac{2}{N}(1-\gamma s) \right] .
\end{eqnarray}
The stochastic differential equation including noise is then obtained as 
\begin{equation}
\frac{1}{\alpha} \frac{ds}{dt} = -N \gamma s^2 -2s + \gamma (2+N) + \sqrt{ \frac{2}{\alpha}\left[(1+s)(1-s) + \frac{2}{N} (1- \gamma s)\right] } \xi(t) .
\end{equation}
A straightforward generalization to the multi-site case gives the following evolution equations
\begin{eqnarray}
\frac{1}{\alpha} \frac{ds_i}{dt} & = -N \gamma_i(\bm{s})  s_i^2 - 2s_i + \gamma_i(\bm{s}) (2+N) \nonumber \\
& + \sqrt{ \frac{2}{\alpha} \left[ (1+s_i)(1-s_i) + \frac{2}{N} (1- \gamma_i(\bm{s}) s_i) \right] } \xi_i(t) .
\label{finalsde}
\end{eqnarray}
The above equation is identical to (\ref{alleqns}) up to the Gaussian noise term.  This allows for the system
to escape local minima in the energy landscape.  The steady state evolutions then approach the correct thermal
equilibrium averages as defined by the Boltzmann distribution. Combined with an annealing schedule, this may 
be used to find the ground state of the Ising Hamiltonian.  As found previously, the due to the factors of $ N $ in 
(\ref{finalsde}) originating from bosonically enhanced cooling, the annealing rates may be made $ N $ times
faster, allowing for an accelerated method of finding the solution to Ising model problems, as claimed in Ref. \cite{byrnes11}.

\section{Learning and Pattern Completion}
\label{secimplementation}

In the context of the neural networks, learning and memory retrieval are typical 
tasks which illustrate their utility. Due to the equivalence of the 
BEC network to a continuous Hopfield model, this implies that such processes should also 
apply to BEC networks. In this section we illustrate the equivalence between the two 
systems by simple examples of Hebbian learning and 
pattern completion.

\subsection{Learning}

The simplest example of unsupervised learning is the Hebbian learning rule \cite{zurada92}. Using
the associations in Table \ref{table1} it is straightforward to write down the corresponding 
rule in the case of the BEC network.  We follow the presentation give in Ref. \cite{zurada92} (sec. 2.5) for
the case of continuous activation functions, since in the BEC system the measured spin is a continuous
quantity. We assume that the BEC network starts with the system shown in Figure \ref{fig1} with the Ising
matrix set to 
\begin{equation}
J_{ij}=0 .
\end{equation}
The learning algorithm then proceeds as follows.  We apply various magnetic field configurations $ \lambda_i^{(n)} $ where $ n $ 
labels the various pattern configurations that the network is exposed to during the learning process. Starting with 
the first field configuration $ n = 1$, we apply this field and wait until the spins reach their equilibrium value, 
which will be given (for the first iteration)
\begin{equation}
s_i = \Phi(\beta \lambda_i^{(1)} )
\label{firstspins}
\end{equation}
We then update the Ising matrix according to
\begin{equation}
J_{ij} \rightarrow J_{ij} + c s_i \lambda_j^{(n)}
\label{hebbian}
\end{equation}
where $ c $ is the learning constant which determies the speed of the learning process.  We then make subsequent applications
of the field $ \lambda_i^{(n)} $, measure the field $ s_i $ in each case, then make the replacement (\ref{hebbian}).  For $ n\ge 2 $, the spins do not simply take the values of (\ref{firstspins}) since the  fields $ B_i $ will in general be non-zero.  The process
is continued until the learning examples are exhausted, or the same set can be recycled.  Other learning algorithms 
may be derived in a similar way using the associations in Table \ref{table1}.

\begin{figure}
\scalebox{0.7}{\includegraphics{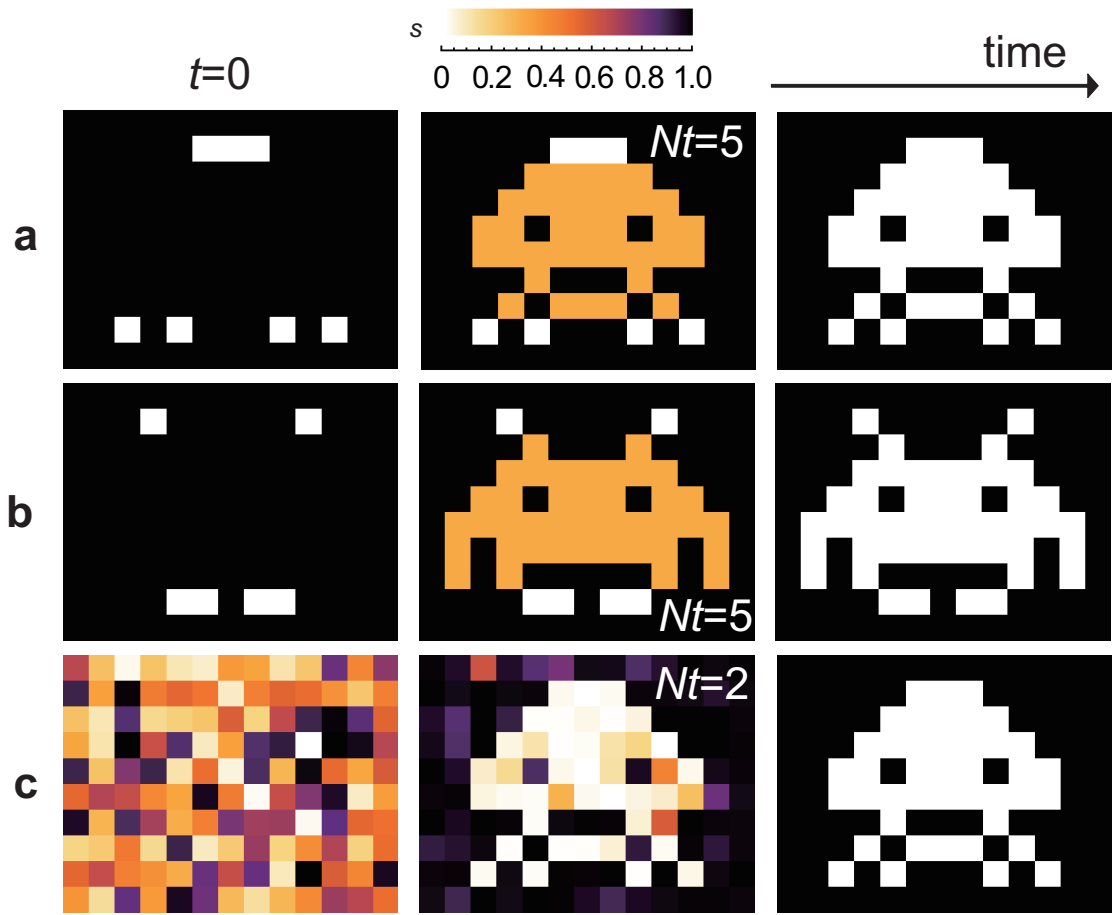}}
\caption{\label{fig2}
Pattern completion for a BEC network prepared with Hebbian learning for the two space invader characters in (a) and (b).  
The BEC network is evolved in time starting from the initial conditions as shown and evolve towards their steady state configurations. 
Parameters used are $ \alpha = 1 $, $ N= 10^4 $, $ k_B T = 0 $.}
\end{figure}

\subsection{Pattern completion}

As a simple example of the use of the BEC network for pattern completion, we test the set of equations given in (\ref{alleqns}) using
an Ising matrix trained using the Hebbian algorithm of the previous section. We numerically evolve a set of $13 \times 10 $ equations
forming a two dimensional grid at zero temperature from the initial conditions as shown in Figure \ref{fig2}.  In Figures \ref{fig2}a and \ref{fig2}b we start from 
fragments of the learned patterns while in Figure \ref{fig2}c we start from a randomly chosen spin configuration.  We see that in all cases 
the spins evolve towards the learned configurations, with the BEC network completing the patterns as desired.  For the random initial configuration, the spins evolve towards whichever configuration happens to be closer to the learned patterns. 

\begin{figure}
\scalebox{0.7}{\includegraphics{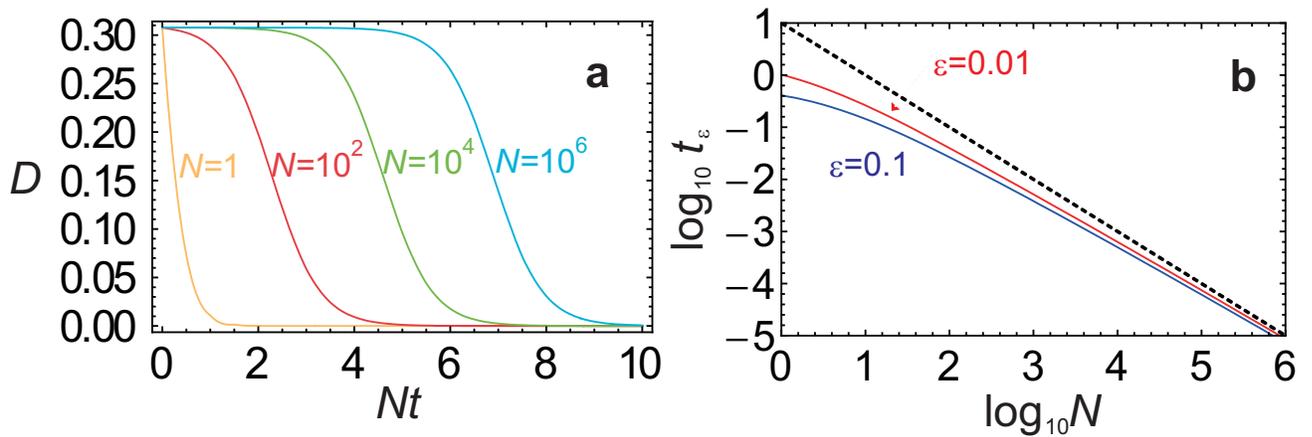}}
\caption{\label{fig3}
(a) Hamming distance $ D$ versus time for the pattern recognition task given in Figure \ref{fig2}a for various boson numbers $N $ as 
shown.  (b) Times $ t_{\varepsilon} $ necessary for reaching a Hamming distance of $ \varepsilon $ as a function of $ N $. Dotted line corresponds to $ 1/N $.}
\end{figure}

The time scaling behaviour is shown in Figure \ref{fig3}.  In Figure \ref{fig3}a we plot the normalized Hamming distance 
\begin{equation}
D= \frac{1}{2M}\sum_i |s_i(t) - s_i^{(n)}|
\end{equation}
between the evolved spin configuration and the learned spin configurations $ s_i^{(n)} $. We see that the general behavior is that the 
time for the pattern completion scales as $ \sim 1/N $, which can again be attributed to bosonic stimulated cooling. There is a logarithmic correction to this behavior, where there is an initial stiffness of the spins to move towards the steady state configuration.  
In Figure \ref{fig3}b we show the scaling of the time to reach a particular Hamming distance $ \varepsilon $ with respect to $ N $.  We
see that for large $ N $ all curves converge to the dominant $ \sim 1/N $ behaviour.  This shows that BEC networks can equally well 
be used to perform tasks such as pattern completion, with the additional benefit of a reduced time in proportion to $ N $.

\section{Summary and conclusions}
\label{secconclusions}

We have analyzed the BEC network proposed in Ref. \cite{byrnes11} in terms of the theory of neural
networks and found that it is equivalent to a stochastic continuous Hopfield model. In contrast to 
the continuous Hopfield model where the overall timescale of the evolution is determined by the 
capacitance within each unit, in the BEC network the timescale is determined via the rate of cooling.
Due to bosonic stimulated cooling, the rate of cooling may be accelerated in proportion to the number
of bosons $ N $ on each site, which in turn accelerates the cooling rate of the entire system. 
The bosonic stimulated cooling makes the time evolution equations (\ref{alleqns}) on a single site 
not precisely the same as its Hopfield model counterpart (\ref{hopfielddyn}), but the difference 
merely gives a modification of the dynamics as the system heads towards equilibrium, the overall
behaviour of the system as a whole remains the same.  In particular, tasks such as pattern 
completion may be performed using the BEC network, in the same way as the Hopfield model. 

In this context, it would appear that using a 
BEC network, rather than a physical implemetation of a Hopfield network, is nothing but a more 
complicated way of implementing what could be 
done equally well by either standard electronics or optical means \cite{rojas96}. 
Specifically, one could imagine using simply Hopfield circuits with small capacitances such that the 
timescale of the circuit is as small as desired. Other variations of optical implementations
of Hopfield models allow for fast operation speeds. While for the zero temperature 
case this may be true, the BEC system does have the advantage that the random fluctuations 
following Boltzmann statistics is already built-in, and do not require additional circuitry 
to simulate. Another possible advantage is that the speedups can be made systematically faster by simply 
by increasing the number of bosons. A possible issue for the physical realization is whether the 
connections between each Ising site require response
times of the order of the cooling time on each site.  Apart from a simple slowdown due to 
bottlenecks in the transmission, such delay times in the information between each site can 
introduce instabilities in the system causing divergent behavior \cite{yan13}.   
We leave as future work whether the proposals in Refs. \cite{utsunomiya11} and \cite{takata12}
can be treated with the same analysis.

\ack

This work is supported by the Transdisciplinary Research Integration Center, Special Coordination Funds for Promoting Science and Technology, Navy/SPAWAR Grant N66001-09-1-2024, Project for Developing Innovation Systems of MEXT, and the JSPS through its FIRST program.

\end{document}